\documentclass[11pt,a4paper,twoside,final]{scrartcl}

\usepackage{a4wide}
\usepackage{amsfonts}
\usepackage{amsmath}
\usepackage{amssymb}
\usepackage{amsthm}
\usepackage{enumerate}
\usepackage{subfig}
\usepackage{algorithm}
\usepackage{algorithmic}
\usepackage{pgfplots}
\pgfplotsset{compat=newest}
\usepackage{graphicx}
\graphicspath{{./files/}}
\usepackage{color}
\usepackage{colortbl}
\usepackage{showkeys}
\usepackage{verbatim}
\usepackage{hyperref}
\usepackage{todonotes}
\usepackage{mathtools}
\usepackage{booktabs}
\usepackage{jlcode}
\usepackage{dsfont}
\usepackage{makecell}

\usepackage{pgfplots}
\pgfplotsset{compat=1.8}
\usepackage{tikz}
\usetikzlibrary{external}
\tikzexternaldisable

\renewcommand{\d}{\,\mathrm{d}}
\newcommand{\e}{\mathrm e}
\renewcommand{\i}{\mathrm i}

\renewcommand{\b}{\boldsymbol} 
\newcommand{\R}{\mathbb R}
\newcommand{\C}{\mathbb C}
\newcommand{\Z}{\mathbb Z}
\newcommand{\N}{\mathbb N}
\newcommand{\T}{\mathbb T}
\renewcommand{\H}{\mathcal H}

\newcommand{\abs}[1]{|#1|}
\newcommand{\norm}[1]{\| #1 \|}

\newtheorem{theorem}{Theorem}[section]
\newtheorem{lemma}[theorem]{Lemma}
\newtheorem{remark}[theorem]{Remark}
\newtheorem{generalisation}[theorem]{Generalisation}
\newtheorem{definition}[theorem]{Definition}
\newtheorem{example}[theorem]{Example}
\newtheorem{corollary}[theorem]{Corollary}
\newtheorem{proposition}[theorem]{Proposition}

\makeatletter
\def\imod#1{\allowbreak\mkern10mu({\operator@font mod}\,\,#1)}
\makeatother

\makeatletter
\renewcommand{\todo}[2][]{\tikzexternaldisable\@todo[#1]{#2}\tikzexternalenable}
\makeatother

\makeatletter 
 
\@addtoreset{algorithm}{chapter} 
\makeatother

\numberwithin{equation}{section}
\numberwithin{table}{section}
\numberwithin{figure}{section}

\newcommand{\bend}{\hspace*{0ex} \hfill \hbox{\vrule height
    1.5ex\vbox{\hrule width 1.4ex \vskip 1.4ex\hrule  width 1.4ex}\vrule
    height 1.5ex}}

\long\def\symbolfootnote[#1]#2{\begingroup
\def\thefootnote{\fnsymbol{footnote}}\footnote[#1]{#2}\endgroup}

\pgfplotsset{
	log x ticks with fixed point/.style={
		xticklabel={
			\pgfkeys{/pgf/fpu=true}
			\pgfmathparse{exp(\tick)}
			\pgfmathprintnumber[fixed relative, precision=3]{\pgfmathresult}
			\pgfkeys{/pgf/fpu=false}
		}
	},
	log y ticks with fixed point/.style={
		yticklabel={
			\pgfkeys{/pgf/fpu=true}
			\pgfmathparse{exp(\tick)}
			\pgfmathprintnumber[fixed relative, precision=3]{\pgfmathresult}
			\pgfkeys{/pgf/fpu=false}
		}
	},
	log x ticks with f point/.style={
		xticklabel={
			\pgfkeys{/pgf/fpu=true}
			\pgfmathparse{exp(\tick)}
			\pgfmathprintnumber[fixed relative, precision=4]{\pgfmathresult}
			\pgfkeys{/pgf/fpu=false}
		}
	}
}

\title{Nonequispaced Fast Fourier Transform (NFFT) Interface for Julia}

\date{}
\author{Michael Schmischke\footnotemark[1]}

\hypersetup{pdfauthor={Michael Schmischke},
          pdftitle={Nonequispaced Fast Fourier Transform (NFFT) Interface for Julia},
          pdfsubject={},
          pdfcreator = {pdflatex},
          plainpages=false,
          pdfstartview=FitH,       
          pdfview=FitH,            
          pdfpagemode=UseOutlines, 
          bookmarksnumbered=true, 
          bookmarksopen=false,     
          bookmarksopenlevel=0,   
          colorlinks=true,       
          linkcolor=black,
          citecolor=black,
          urlcolor=black}
\begin{document}

\maketitle

\begin{abstract}

This report describes the newly added Julia interface to the NFFT3 library. We explain the multidimensional NFFT algorithm and basics of the interface. Furthermore, we go into detail about the different parameters and how to adjust them properly. 

\medskip


\medskip

\end{abstract}
\footnotetext[1]{
  Technische Universit\"at Chemnitz, Faculty of Mathematics, 09107 Chemnitz, Germany\\ michael.schmischke@math.tu-chemnitz.de
}

\medskip

\section{Introduction}

The nonequispaced fast Fourier transform \cite{KeKuPo06} (NFFT or NUFFT) overcomes one of the main shortcomings of the FFT - the need for an equispaced sampling grid. Considering a $d$-dimensional trigonometric polynomial 
\begin{equation}
	f(\b x) \coloneqq \sum_{\b k \in I_{\b N}} \hat{f}_{\b k} \e^{-2\pi\i\b{k}\b{x}}
\end{equation}
with an index set $I_{\b N} \coloneqq \{ \b k \in \Z^d: -\frac{N_i}{2}\leq \b{k}_i \leq\frac{N_i}{2}-1, i=0,\dots,d-1 \}$ where $\b N \in 2\N^d$ is the multibandlimit, the NDFT is its evaluation at $M \in \N$ nonequispaced points $\b{x}_j \in \T^d$ for $j = 0, 1, \dots, M$,
\begin{equation}\label{eq:NDFT}
f(\b{x}_j) =\sum_{\b k \in I_{\b N}} \hat{f}_{\b k} \e^{-2\pi\i\b{k}\cdot\b{x}_j},
\end{equation} 
with given coefficients $\hat{f}_{\b k} \in \C$ where we identify the smooth manifold of the torus $\T$ with $[-1/2, 1/2)$. The NFFT is an algorithm for the fast evaluation of the sums \eqref{eq:NDFT} and the adjoint problem, the fast evaluation of
\begin{equation}
	\hat{h}_{\b k} = \sum_{j = 0}^{M-1} f_j \e^{2\pi\i\b{k}\b{x}_j}, \b k \in I_{\b N}
\end{equation}
for given coefficients $f_j \in \C$. The available NFFT3 library \cite{nfft3} provides C routines for the NFFT, applications such as the fast evaluation of sums
\begin{equation}
g(\b{y}_j) \coloneqq \sum_{k=1}^{N}\alpha_k K(\norm{\b{y}_j-\b{x}_k}_2), j = 1, \dots, M
\end{equation}
for given coefficients $\alpha_k \in \C$, nodes $\b{x}_k,\b{y}_j \in \R^d$  and a radial kernel function $K: [0,\infty) \to [0,\infty)$, and generalizations such as the NNFFT for nonequispaced nodes in time and frequency domain. The major NFFT3 release included interfaces for the numerical computing environments MATLAB and OCTAVE, expanding the user base.  

In the past years, a new dynamic programming language called Julia, see \cite{Bezanson2017}, caught interest in the field of numerical computing. The language is designed around the idea of high performance and flexibility with a lot of available packages for different purposes. The growing community around Julia has led to the idea of further expanding the NFFT3's user base by adding a corresponding interface to the library. 

After discussing the basic idea of the NFFT in Section \ref{sec:nfft}, we go into detail on the interface in Section \ref{sec:interface} starting with a manual on usage and parameters. Furthermore, we discuss the basic structure and Julia functions used. Section \ref{sec:examples} contains multiple examples.

\clearpage

\section{NFFT Algorithm}\label{sec:nfft}

In this section, we give a brief description of the NFFT algorithm based on \cite{KeKuPo09} and \cite[Chapter 7]{PlPoStTa18}. This will be the basis for the explanation of parameter choices later on.

We consider the evaluation of the $d$-dimensional trigonometric polynomial 
\begin{align}\label{eq:trigpoly}
f: \T^d \to \C, \b x \mapsto \sum_{\b k \in I_{\b N}} \hat{f}_{\b k} \e^{-2\pi\i{\b k}\cdot{\b x}}
\end{align}
with multibandlimit $\b N \in 2\N^d$ and index set
\begin{equation}\label{eq:index_set}
I_{\b N} \coloneqq \{ \b k \in \Z^d: -N_i/2\leq k_i \leq{N_i}/{2}-1 \,\,\forall i \in \{1,2,\dots,d\} \}.
\end{equation}
The first approximation is a linear combination of a shifted periodized window function $\tilde{\varphi}$
\begin{equation}\label{eq:gl}
s_1(\b x) = \sum_{\b\ell \in I_{\b n}} g_{\b\ell} \tilde{\varphi}\left(\b x - \frac{1}{\b n}\odot \b\ell\right)
\end{equation}
where $\frac{1}{\b n}$ is the elementwise inversion of the vector $\b n$. We choose an oversampling vector $\b\sigma > 1$ componentwise and obtain the index set by
\begin{equation}
\b n \coloneqq \b\sigma \odot \b N.
\end{equation}
Here, $\odot$ denotes the componentwise product. Note that one could skip the choice of $\b\sigma$ entirely and directly choose $n_i > N_i$ for $i=1,2,\dots,d$. The standard choice in the C library is at the moment \[ n_i = 2^{\lceil \log_2 N_i \rceil + 1}, i=1,2,\dots,d. \] 

If the window function $\varphi: \R^d \to \R$ has a one-periodization 
\begin{equation}\label{eq:window}
\tilde{\varphi}(\b x) = \sum_{\b r \in \Z^d} \varphi(\b x+\b r) 
\end{equation} with a uniformly convergent Fourier series, we use the Poisson summation formula and obtain the Fourier coefficients 
\begin{equation}
c_{\b k}(\tilde{\varphi}) = \hat{\varphi}(\b k).
\end{equation}
Here, $\hat{\varphi}$ is the Fourier transform of $\varphi$ defined by
\begin{equation}
\hat{\varphi}(\b k) = \int_{\T^d} \varphi(\b x) \e^{-2\pi\i\b k\cdot\b x} \d \b x.
\end{equation}
Replacing $\tilde{\varphi}$ by its Fourier series and splitting the sum in \eqref{eq:gl} yields
\begin{align}
s_1(\b x) &= \sum_{\b\ell \in I_{\b n}} g_{\b\ell} \sum_{\b k \in \Z^d} c_{\b k}(\tilde{\varphi}) \e^{-2\pi\i \b k \cdot \left(\b x - \frac{1}{\b n}\odot\b\ell\right) } \nonumber \\ 
& = \sum_{\b k \in \Z^d} c_{\b k}(\tilde{\varphi}) \underbrace{\left(\sum_{\b\ell \in I_{\b n}} g_{\b\ell} \e^{2\pi\i\frac{1}{\b n}\odot(\b k\cdot\b\ell) }\right)}_{\eqqcolon \hat{g}_{\b k}}  \e^{-2\pi\i \b k \cdot \b x } \nonumber \\ 
&= \sum_{\b k \in I_{\b n}} c_{\b k}(\tilde{\varphi}) \hat{g}_{\b k} \e^{-2\pi\i\b k \cdot \b x } +\sum_{\b r \in \Z^d\setminus\{\b 0\}} \sum_{\b k \in I_{\b n}} c_{\b k}(\tilde{\varphi}) \hat{g}_{\b k} \e^{-2\pi\i (\b k+\b n\odot\b r)\cdot\b x }. \label{eq:s}
\end{align}
Furthermore, a comparison of \eqref{eq:s} and \eqref{eq:trigpoly} suggests the choice
\begin{equation}
\hat{g}_{\b k} = \begin{cases} \frac{\hat{f}_{\b k}}{c_{\b k}(\tilde{\varphi})} \quad &\text{for } \b k \in I_{\b N} \\ 0 &\text{for } \b k \in I_{\b n}\setminus I_{\b N}\end{cases}.
\end{equation}
Now, we are able to compute $g_{\b\ell}$ by an FFT of size $n_1 \times n_2 \times \cdots \times n_d$. The error of the first approximation $f \approx s_1$ is called aliasing error. For further analysis, we refer to \cite[Chapter 7]{PlPoStTa18}.

We obtain an approximation $\psi$ for $\varphi$ if the window function is well localized in the spatial domain by the truncation
\begin{equation}\label{eq:window_cutoff}
\psi(\b x) \coloneqq \varphi(\b x) \mathds{1}_{\times_{i = 1}^d [-m/n_i,m/n_i]}(\b x)
\end{equation} 
with a chosen window size $m \ll \min_{i\in\{1,2,\dots,d\}}\{n_i\}, m \in \N$. Following the same scheme as above, we can use the periodization \[ \tilde{\psi}(\b x) = \sum_{\b r \in \Z^d} \psi(\b x+\b r) \] and look at the linear combination 
\begin{equation}
s(\b{x}_j) \coloneqq \sum_{\b\ell \in I_{\b n}} g_{\b\ell} \tilde{\psi}\left(\b{x}_j-\frac{1}{\b n}\odot\b\ell\right).
\end{equation} 
The following calculations show that $s \approx s_1$
\begin{align}
s(\b{x}_j) &= \sum_{\b \ell \in I_{\b n}} g_{\b \ell} \sum_{\b r \in \Z^d} \psi\left(\b{x}_j-\frac{1}{\b n}\odot\b\ell+\b r\right) \nonumber \\
&= \sum_{\b \ell \in I_{\b n}} g_{\b\ell} \sum_{\b r \in \Z^d} \varphi\left(\b{x}_j-\frac{1}{\b n}\odot\b\ell+\b r\right) \mathds{1}_{\times_{i = 1}^d [-m/n_i,m/n_i]}(\b{x}_j) \nonumber \\
&= \sum_{\ell \in I_n} g_{\b \ell} \mathds{1}_{\times_{i = 1}^d [-m/n_i,m/n_i]}(\b{x}_j) \tilde{\varphi}\left(\b{x}_j-\frac{1}{\b n}\odot\b\ell\right) \nonumber \\
&= \sum_{\b \ell \in I_{\b n,m}(\b{x}_j)} g_{\b\ell} \tilde{\varphi}\left(\b{x}_j-\frac{1}{\b n}\odot\b\ell\right) \label{eq:s_approx}
\end{align}
with the index set 
\begin{equation}\label{eq:Inm}
I_{\b n,m}(\b{x}_j) = \{ \b\ell \in I_{\b n} : \b n \odot \b{x}_j-m\b{1} \leq \b\ell \leq \b n \odot \b{x}_j+m\b{1} \}
\end{equation} 
for a fixed node $\b{x}_j$. This is motivated by \[ -\frac{m}{n_i} \leq \left(\b{x}_j\right)_i \leq \frac{m}{n_i} \] in order to ensure that $\b{x}_j$ is within the support. This second approximation error is called the truncation error. Summarizing, we have $f \approx s_1 \approx s$.

This covers the basic idea of the algorithm, which is given in pseudo-code in Algorithm \ref{alg:nfft}. Using the transposed index set 
\begin{equation}
I_{\b n,m}^\top(\b\ell) = \{ j=0,1,\dots,M-1 : \b\ell-m\b{1} \leq \b n \odot \b{x}_j \leq \b\ell+m\b{1} \},
\end{equation}
we obtain the adjoint NFFT algorithm, given in pseudo-code in Algorithm \ref{alg:adjnfft}. Further information regarding specific window functions with error bounds and applications is given in \cite[Chapter 7.2]{PlPoStTa18}.

\begin{algorithm}[ht]
	\vspace{2mm}
	\begin{tabular}{ l l l }
		\textbf{Input:} & $d \in \N$ & dimension \\
		& $\b N \in 2\N^d$ & degree of polynomial \\
		& $M \in \N$ & number of nodes \\
		& $\b{x}_j \in \T^{d}, j = 0,\dots,M-1$ & nodes \\
		& $\hat{f}_{\b k} \in \C, \b k \in I_{\b N}$ & Fourier coefficients \\
	\end{tabular}
	\begin{algorithmic}[1]
		\FOR{$\b k \in I_{\b N}$}
		\STATE{$\hat{g}_{\b k} \leftarrow \abs{I_{\b n}}^{-1}\cdot\hat{f}_{\b k}/c_{\b k}(\tilde{\varphi})$}
		\ENDFOR
		\FOR{$\b \ell \in I_{\b n}$}
		\STATE{$\hat{g}_{\b \ell} \leftarrow \sum_{\b k \in I_{\b N}} \hat{g}_{\b k}\e^{-2\pi\i\b k\cdot\left( \frac{1}{\b n}\odot \b \ell\right)} \hspace*{\fill} \vartriangleleft $ $d$-variate FFT }
		\ENDFOR
		\FOR{$j=0,1,\dots,M-1$}
		\STATE{$f_j \leftarrow \sum_{\b \ell \in I_{\b n,m}(\b{x}_j)} g_{\b\ell}\tilde{\varphi}\left(\b{x}_j-\frac{1}{\b n} \odot \b \ell\right)$}
		\ENDFOR
	\end{algorithmic}
	\begin{tabular}{ l l l }
		\textbf{Output:} & $f_j \in \C, j=0,1,\dots,M-1$ & approximate function values \\
	\end{tabular}
	\begin{tabular}{ l l }
		\textbf{Arithmetic cost:} & $\abs{I_{\b N}}+\abs{I_{\b n}}\log\abs{I_{\b n}}+2(2m+1)^dM$ + eval. of window function \\
	\end{tabular}
	\caption{Nonequispaced fast Fourier transform (NFFT)}
	\label{alg:nfft}
\end{algorithm}

\begin{algorithm}[ht]
	\vspace{2mm}
	\begin{tabular}{ l l l }
		\textbf{Input:} & $d \in \N$ & dimension \\
		& $\b N \in 2\N^d$ & degree of polynomial \\
		& $M \in \N$ & number of nodes \\
		& $\b{x}_j \in \T^{d}, j = 0,\dots,M-1$ & nodes \\
		& $\hat{f}_{\b k} \in \C, \b k \in I_{\b N}$ & Fourier coefficients \\
	\end{tabular}
	\begin{algorithmic}[1]
		\FOR{$\b \ell \in I_{\b n}$}
		\STATE{$g_{\b\ell} \leftarrow \sum_{j \ell \in I_{\b n,m}^{\top}(\b\ell)} f_{j}\tilde{\varphi}\left(\b{x}_j-\frac{1}{\b n} \odot \b \ell\right)$}
		\ENDFOR
		\FOR{$\b k \in I_{\b N}$}
		\STATE{$\hat{g}_{\b k} \leftarrow \sum_{\b \ell \in I_{\b n}} \hat{g}_{\b \ell}\e^{2\pi\i\b k\cdot\left( \frac{1}{\b n}\odot \b \ell\right)} \hspace*{\fill} \vartriangleleft $ $d$-variate inverse FFT }
		\ENDFOR
		\FOR{$\b k \in I_{\b N}$}
		\STATE{$\hat{h}_{\b k} \leftarrow \abs{I_{\b n}}^{-1}\cdot\hat{g}_{\b k}/c_{\b k}(\tilde{\varphi})$}
		\ENDFOR
	\end{algorithmic}
	\begin{tabular}{ l l l }
		\textbf{Output:} & $\hat{h}_{\b k} \in \C, \b k \in I_{\b N}$ & approximate coefficients \\
	\end{tabular} \\
	\begin{tabular}{ l l }
		\textbf{Arithmetic cost:} & $\abs{I_{\b N}}+\abs{I_{\b n}}\log\abs{I_{\b n}}+2(2m+1)^dM$ + eval. of window function \\
	\end{tabular}
	\caption{Adjoint nonequispaced fast Fourier transform (NFFT)}
	\label{alg:adjnfft}
\end{algorithm}

We consider the relative errors
\begin{align}
E_2 &\coloneqq \frac{\norm{\b f-\tilde{\b f}}_2}{\norm{\b f}_2} \label{eq:E2} \\
E_\infty &\coloneqq \frac{\norm{\b f-\tilde{\b f}}_\infty}{\norm{\hat{\b f}}_1} \label{eq:EI}
\end{align}
with $\b f = (f(\b{x}_j))_{j=0}^{M-1}$, $\tilde{\b f} = (s(\b{x}_j))_{j=0}^{M-1}$ and $\hat{\b f} = (\hat{f}_{\b k})_{\b k \in I_{\b N}}$ in order to check the quality of the approximation with fixed polynomial $f$.

Furthermore, we want to note that the NFFT can be parallelized which is realized through OpenMP in the NFFT3 library, see \cite{Vo12}.

\clearpage

\section{Documentation and Notes on Implementation}\label{sec:interface}

In this section, we go into detail about the usage of the interface. Specifically, we explain the parameter choices and how they may impact computation times and accuracy of the results. Furthermore, Section \ref{sec:impl} contains a brief introduction to the basics of the implementation.

\subsection{Usage and Parameters}\label{sec:usage}

The NFFT plan structure is the core of the NFFT Julia interface and has the parameters listed in Table \ref{tab:params}. We go into detail about every parameter, its standard choice and how to tweak it for more speed or accuracy.

\begin{table}[ht]
	\begin{center}
		\begin{tabular}{l|l|l}
			\toprule
			parameter & Julia type & description \\ \midrule
			$\b N \in 2\N^d$ & NTuple\{D,Int32\} & degree of polynomial $f$ given in \eqref{eq:trigpoly} \\
			$M \in \N$ & Int32 & number of nodes $\b{x}_j \in \T^d$ \\
			$\b n \in 2\N^d$ & NTuple\{D,Int32\} & $n_i > N_i$ FFT length \\
			$m \in \N$ & Int32 & window size \\
			$\mathrm{f1} \in \N$ & UInt32 & NFFT flags \\
			$\mathrm{f2} \in \N$ & UInt32 & FFTW flags \\
			\bottomrule
		\end{tabular}
		\caption{Fields of the NFFT structure.}
		\label{tab:params}
	\end{center}
\end{table}

The first parameter $\b N \in 2\N^d$ is the multibandlimit of the trigonometric polynomial $f$, see \eqref{eq:trigpoly}. It is required to consist of even numbers and is pre-determined by the problem and therefore a mandatory parameter. This parameter has a direct influence on the computation time since it determines the number of coefficients $\hat{g}_{\b k} \in \C, \b k \in I_{\b N}$. Furthermore, it is a lower bound for $\b n$ which in turn determines the number of $g_{\b l}$ in \eqref{eq:s_approx} and therefore the length of the FFT. 

The parameter $M$ is simply the number of given nodes $\b{x}_j$ where one wants to evaluate $f$. This translates to the number of times one has to compute the sum \eqref{eq:s_approx}. Since $M$ is also determined by the problem, it is mandatory.

At this point, we are able to initialize an NFFT plan. The module provides a constructor \begin{lstlisting}
Plan(N::NTuple{D,Integer},M::Integer)
\end{lstlisting} for initialization with no further parameter choices. The interface will use standard values for the other parameters that we will explain later on. Any integer value that is not of type Int32 will be converted accordingly.

Starting with the FFT length $\b n \in 2\N^d$, every additional parameter is optional. The current standard choice for $\b n$ is \[ n_i = 2^{\lceil \log_2 N_i \rceil + 1}, i=1,2,\dots,d, \] which corresponds to the NFFT3 library standard choice and ensures good accuracy. In general, one can use every even number larger than $n_i$, e.g. \[ N_i < n_i \leq 2 N_i \] with higher computation time for bigger values. Note that this statement has to be relativized because of the small impact of the FFT time compared to the entire NFFT, see Section \ref{sec:fftlen}.

The window size $m$ determines the size of the support of the approximation $\psi$ for the window function $\varphi$, see \eqref{eq:window_cutoff}. The standard choice is \[ m = 8. \] In theory, one could choose any positive integer as the window size as long as we have $m \ll \min\{n_i\}$. The accuracy of the result decreases with smaller $m$ and increases with larger $m$. Section \ref{sec:window} contains tests for one example and many window sizes. One can achieve machine precision with $m=8$ as we will see. Note that those are statements for the standard Kaiser-Bessel window and results for $m$ vary based on the window function.

In Table \ref{tab:nfftflags}, we list every NFFT flag that is defined in Julia. In order to set flags, one has to use the bitwise or operation. The standard flags are defined as \begin{lstlisting}[breaklines=true]
f1_default_1d = PRE_PHI_HUT | PRE_PSI | MALLOC_X | MALLOC_F_HAT | MALLOC_F | FFTW_INIT | FFT_OUT_OF_PLACE
\end{lstlisting} for a 1-dimensional NFFT and \begin{lstlisting}[breaklines=true]
f1_default = PRE_PHI_HUT | PRE_PSI | MALLOC_X | MALLOC_F_HAT | MALLOC_F | FFTW_INIT | FFT_OUT_OF_PLACE | NFFT_SORT_NODES | NFFT_OMP_BLOCKWISE_ADJOINT
\end{lstlisting} for $d>1$.

\begin{table}[ht]
	\begin{center}
		\begin{tabular}{l|p{6cm}} \toprule
			flag name & description \\ \midrule
			PRE\_PHI\_HUT & precompute and store values $\hat{\varphi}(\b k)$ of the Fourier transform of the window function $\hat\varphi$ \\ \midrule
			FG\_PSI & fast Gaussian gridding (see \cite{KuPo06}) \\ \midrule
			PRE\_LIN\_PSI & linear interpolation of the window function from a lookup table (see \cite{KuPo06}) \\ \midrule
			PRE\_FG\_PSI & fast Gaussian gridding (see \cite{KuPo06}) \\ \midrule
			PRE\_PSI & precomputation based on tensor product structure of the window function (see \cite{KuPo06}) \\ \midrule
			PRE\_FULL\_PSI & calculate and store all values $\tilde{\psi}\left(\b{x}_j-\frac{1}{\b n}\odot\b\ell\right)$, see \eqref{eq:s_approx} \vspace{4pt} \\ \midrule
			MALLOC\_X* & allocate memory for nodes $\b{x}_j$ \\  \midrule
			MALLOC\_F\_HAT* & allocate memory for coefficients $\hat{f}_{\b k}$ \\ \midrule
			MALLOC\_F* & allocate memory for approximate function values ${f}_{j}$ \\ \midrule
			FFT\_OUT\_OF\_PLACE & FFTW uses disjoint input/output vectors \\ \midrule
			FFTW\_INIT* & initialize FFTW plan \\ \midrule
			NFFT\_SORT\_NODES & internal sorting of the nodes $\b{x}_j$ that may increase performance (see \cite{Vo12}) \\ \midrule
			NFFT\_OMP\_BLOCKWISE\_ADJOINT & blockwise calculation for adjoint NFFT in the case of OpenMP, see \cite{Vo12} \\ \bottomrule
		\end{tabular}
		\caption{NFFT flags. Flags marked with a * are always set when using the Julia interface.}
		\label{tab:nfftflags}
	\end{center}
\end{table}

As for the NFFT, one can set flags for the FFTW \cite{fftw} which is used to compute the coefficients $g_{\b \ell}$, see \eqref{eq:gl}. All available flags and a brief description are listed in Table \ref{tab:fttwflags}. We are using \begin{lstlisting}[breaklines=true]
f2_default = FFTW_ESTIMATE | FFTW_DESTROY_INPUT
\end{lstlisting} as the standard choice.

\begin{table}[ht]
	\begin{center}
		\begin{tabular}{l|p{8cm}} \toprule
			flag name & description \\ \midrule
			FFTW\_MEASURE &  find optimal plan by executing several FFTs and compare times \\ \midrule
			FFTW\_PATIENT & behaves like FFTW\_MEASURE with a wider range of tests \\ \midrule
			FFTW\_EXHAUSTIVE & behaves like FFTW\_PATIENT with an even wider range of tests \\ \midrule
			FFTW\_WISDOM\_ONLY & a plan is only created if wisdom from tests is available \\ \midrule
			FFTW\_ESTIMATE & use simple heuristic instead of measurements to pick a plan \\ \midrule
			FFTW\_DESTROY\_INPUT & an out-of-place transform is allowed to overwrite the input array with arbitrary data \\ \midrule
			FFTW\_UNALIGNED & the algorithm may not impose any unusual alignment requirements on the input/output arrays (not necessary in most context) \\ \midrule
			FFTW\_CONSERVE\_MEMORY & conserving memory \\ \midrule
			FFTW\_PRESERVE\_INPUT & input vector is preserved and unchanged \\ \bottomrule
		\end{tabular}
		\caption{FFTW flags, see \cite{fftw}.}
		\label{tab:fttwflags}
	\end{center}
\end{table}

In order to set any of the additional parameters, one has to use the advanced constructor \begin{lstlisting}[breaklines=true]
Plan(N::NTuple{D,Integer},M::Integer,n::NTuple{D,Integer},m::Integer=Int32(6),f1::UInt32=(D > 1 ? f1_default : f1_default_1d),f2::UInt32=f2_default), 
\end{lstlisting} but it is possible to pass on the specification of parameters that appear later in the function head. Section \ref{sec:examples} offers some examples.

\subsection{Implementation}\label{sec:impl}
The intention for the Julia interface is to use the C library to the most possible extent. In contrast to MATLAB that relies on MEX files as the only way to use existing C subroutines, Julia allows direct calling of C functions within shared libraries. More advantages are the direct passing of variables and pointers between C and Julia and further allowing us to access the allocated memory of the other language. Taking this into account, we have good prerequisites for a lightweight Julia interface.

Now, we discuss the basic idea of the Julia code. At the core, we are working with\begin{lstlisting}[breaklines=true]
ccall((function, library), returntype, (argtype1, ...), argvalue1, ...)
\end{lstlisting} in order to call a function in a shared library. It is possible to pass Julia variables to C directly such as integers or doubles, however arrays have to be passed as pointers, consistent with what we would expect in C. The same is also true for the return value of the respective function. In addition, one has to convert the type of certain variables according to a matching table.

In order to match those specific needs and reduce the number of times we have to use \verb|ccall|, we are using the NFFT library and expand it with new functions tailored to be called by Julia. The new library is called libnfftjulia.c. Another reason for this separate library is that although one can compile custom C code to access fields of a C struct, this would bring unnecessary computation cost.

Consistent with the goal of a lightweight interface, we want to use as little additional memory as possible. The Julia structure for an NFFT plan only holds basic parameters while we use the memory allocated by C to store the node set, values and coefficients. For further processing of those by users, we are storing the C pointer to the containing array (therefore eliminating the need for an additional \verb|ccall|) after the transformation or adjoint transformation respectively.

\clearpage

\section{Examples and Time Comparison}\label{sec:examples}

In this chapter, we look at examples for using the NFFT Julia interface and how parameter choices effect accuracy and computation time. In order to have comparable results, we solely tested the three trigonometric polynomials 
\begin{equation}\label{eq:trigpolys}
f_d: \T^d \to \C, \b x \mapsto \sum_{\b k \in I_{d}} \frac{1}{1+\norm{\b k}_2} \e^{-2\pi\i{\b k}\cdot{\b x}}, d = 1,2,3,
\end{equation}
with multibandlimits $N_1 = 512$, $\b{N}_2 = (128,128)$, $\b{N}_3 = (32,32,32)$ and corresponding index sets $I_1 \coloneqq I_{N_1}$, $I_2 = I_{\b{N}_2}$, $I_3 = I_{\b{N}_3}$ given by \eqref{eq:index_set}. We use three sets of random (equally distributed) nodes $\b{x}_j, j=0,1,\dots,M_d-1$ with $M_1 = 2\cdot 512$, $M_2 = 2\cdot {128}^2$, $M_3 = 2\cdot {32}^3$, one for each polynomial $f_1$, $f_2$ and $f_3$. Additionally, we are using the Kaiser-Bessel window, see \cite[Chapter 7.2]{PlPoStTa18}, for $\varphi$ in \eqref{eq:window}.

In Section \ref{sec:nfft3lib} we explain the usage of the NFFT3 library especially choices for the configuration such as OpenMP. Section \ref{sec:parfree} offers examples without additional parameters and Section \ref{sec:pars} contains comparisons between FFT length $\b n \in 2\N^d$ and window size $m \in \N$. We will also look at how the number of OpenMP threads impacts computation times in Section \ref{sec:openmp}. 

All tests in this chapter were done on a computer with four Intel(R) Xeon(R) E7-4880 v2 processors with 15 cores and 2.50GHz.

\subsection{NFFT3 Library}\label{sec:nfft3lib}

In order to work with the NFFT in Julia, one needs to download and compile (or install) the NFFT3 library. The NFFT homepage \href{http://tu-chemnitz.de/~potts/nfft}{tu-chemnitz.de/\textasciitilde potts/nfft} offers various stable packages to download, but one could also clone the Github repository \href{https://github.com/NFFT/nfft}{github.com/NFFT/nfft}. 

The README offers detailed explanations on how to compile or install the library, nevertheless we want to briefly describe the process. 
\begin{lstlisting}
./bootstrap.sh 
./configure --enable-julia
make
\end{lstlisting}
Here, bootstrap.sh is a shell script that runs libtoolize and autoreconf since we are working from a source repository. Otherwise, this is not necessary. We are running the configure script with the option to enable Julia. This makes sure that the Julia library is compiled. One can use
\begin{lstlisting}
./configure --help
\end{lstlisting}
to find all options offered by the configure script, some of which we explain below.  

In order to enable OpenMP support (see \cite{Vo12}) one can use
\begin{lstlisting}
./configure --enable-openmp --enable-julia
\end{lstlisting}
and set the desired threads via the shell command \verb|export OMP_NUM_THREADS=N| ($N$ should be replaced with the number of threads). 

In order to choose a different window function $\varphi$ (see \ref{eq:window}) one can use 
\begin{lstlisting}
./configure --with-window=WINDOW --enable-julia
\end{lstlisting}
where \verb|WINDOW| has to be replaced by \verb|kaiserbessel|, \verb|gaussian|, \verb|bspline| or \verb|sinc|. Note that the standard Kaiser-Bessel window is best in most cases. For additional information on window functions we refer to \cite[Chapter 7.2]{PlPoStTa18}.

\subsection{Parameter-free Initialization}\label{sec:parfree}

First, we want to present Julia code examples for the evaluation of the three trigonometric polynomials $f_1$, $f_2$ and $f_3$. We won't set any of the additional parameters $\b n$, $m$, $f1$ and $f2$ and therefore implicitly use the standard values as designated in Section \ref{sec:usage}. The one dimensional example follows below while the examples for $d = 2$ and $d = 3$ can be found in \cite{Sc18julia}.

\begin{lstlisting}[numbers=left]
using NFFT
using LinearAlgebra
using BenchmarkTools

# bandwidth
N = 512

# number of nodes
M = 2*N

#################################
###            NFFT           ###
#################################

# create plan without parameters
p = Plan((N,),M)

# generate random nodes
A = rand(M).-0.5

# set nodes with precomputations
# measure time with 10 samples
@benchmark p.x = A samples=10

# generate Fourier coefficients
fhat = [ 1/(1+norm(k)) for k in -N/2:N/2-1 ]

# set Fourier coefficients
p.fhat = complex(fhat)

# actual transform
# measure time with 10 samples
@benchmark NFFT.trafo(p) samples=10

# get approximate function values
f2 = p.f

#################################
###     error calculation     ###
#################################

# define Fourier matrix
F = [ exp(-2*pi*im*x*k) for x in A, k in -N/2:N/2-1 ]

# multiply Fourier matrix with vector of Fourier coefficients
# to obtain function values
f1 = F*fhat

ev = f1-f2
E_2 = norm(ev)/norm(f1)
E_infty = norm(ev,Inf)/norm(fhat,1)

println( "E_2 error:" )
println( E_2 )
println( "E_infty error:" )
println( E_infty )
\end{lstlisting}

From line 38 onward, we calculate the Fourier matrix manually and print the errors $E_2$ and $E_\infty$ as defined in \eqref{eq:E2} and \eqref{eq:EI} respectively. Table \ref{tab:std} shows the resulting errors and computation times for all three polynomials.

\begin{table}[ht!]
	\begin{center}
		\begin{tabular}{c|cccc} \toprule
			dimension & $E_2$ & $E_\infty$ & \makecell{precomputation \\ time} & \makecell{transformation \\ time}  \\ \midrule
			d = 1  & 2.85$\e$-15 & 2.45$\e$-15 & 0.69 ms & 0.06 ms \\
			d = 2  & 8.81$\e$-15 & 6.43$\e$-15 & 45.59 ms & 21.31 ms \\
			d = 3  & 1.06$\e$-14 & 6.86$\e$-15 & 138.31 ms & 896.82 ms \\ \bottomrule
		\end{tabular}
		\caption{Performance results of the NFFT with the three polynomials $f_1$, $f_2$ and $f_3$. All times are the mean of 10 samples.}
		\label{tab:std}
	\end{center}
\end{table}

\subsection{Parameter Comparison}\label{sec:pars}

After looking at the results with standard parameters, we now want to compare the parameter choices for the trigonometric polynomials $f_1$, $f_2$ and $f_3$. We start by looking at the FFT length $\b n$ in Section \ref{sec:fftlen} and go on to the window size $m$ in Section \ref{sec:window}.

The Julia code for a one dimensional NFFT with chosen parameters $\b n$ and $m$ can be found below. 

\begin{lstlisting}[numbers=left]
using NFFT
using LinearAlgebra
using BenchmarkTools

# bandwidth
N = 512

# number of nodes
M = 2*N

# FFT length
n = 2*N

# window size
m = 4

# standard flags for demonstration purposes
f1 = NFFT.PRE_PHI_HUT | NFFT.PRE_PSI | NFFT.FFT_OUT_OF_PLACE
f2 = NFFT.FFTW_ESTIMATE | NFFT.FFTW_DESTROY_INPUT

#################################
###            NFFT           ###
#################################

# create plan with parameters
p = Plan((N,),M,(n,),m,f1,f2)

# generate random nodes
A = rand(M).-0.5

# set nodes with precomputations
# measure time with 10 samples
@benchmark p.x = A samples=10

# generate Fourier coefficients
fhat = [ 1/(1+norm(k)) for k in -N/2:N/2-1 ]

# set Fourier coefficients
p.fhat = complex(fhat)

# actual transform
# measure time with 10 samples
@benchmark NFFT.trafo(p) samples=10

# get approximate function values
f2 = p.f

#################################
###     error calculation     ###
#################################

# define Fourier matrix
F = [ exp(-2*pi*im*x*k) for x in A, k in -N/2:N/2-1 ]

# multiply Fourier matrix with vector of Fourier coefficients
# to obtain function values
f1 = F*fhat

ev = f1-f2
E_2 = norm(ev)/norm(f1)
E_infty = norm(ev,Inf)/norm(fhat,1)

println( "E_2 error:" )
println( E_2 )
println( "E_infty error:" )
println( E_infty )
\end{lstlisting}

\subsubsection{FFT Length $\b n$}\label{sec:fftlen}

As mentioned in Section \ref{sec:usage}, the standard choice for the FFT length is \[ n_i = 2^{\lceil \log_2 N_i \rceil + 1}, i=1,2,\dots,d. \] In the case of our choices for $\b N$, this is the same as $n_i = 2 N_i$. In order to compare the accuracy, we test values between $\b N$ and $4\b N$ so that they are fulfilling the requirement ${n}_i > N_i$.
\begin{figure}[ht!]
	\subfloat[error $E_2$]{
		\begin{tikzpicture}
		\begin{loglogaxis}[enlargelimits=false,xmin=500,xmax=2100,ymin=1e-15,ymax=2e-3,height=0.33\textwidth, width=0.43\textwidth, grid=major, xlabel={$n_i$}, ylabel={$E_2$},
		xtick={514,1024,2048}, log x ticks with f point, 
		ytick={1e-3,1e-10,1e-14},
		legend style={at={(0.5,1.07)}, anchor=south,legend columns=3,legend cell align=left, font=\small, 
		},]
		\addplot[blue,mark=o,mark size=2] coordinates {
			(514,0.0004888148623492772) (614,4.874768889496525e-11) (716,1.1661119473999575e-13) (820,4.532895988485328e-15) (922,2.7338339701208107e-15) (1024,2.8505654409202513e-15) (1126,3.119985552570025e-15) (1228,2.742010266536845e-15) (1332,3.168701404921872e-15) (1434,3.4588911420417695e-15) (1536,3.2017652522838687e-15) (1638,4.132823912806892e-15) (1740,2.924406685755084e-15) (1844,3.901049043279157e-15) (1946,3.3426020422346157e-15) (2048,3.179148445558315e-15)
		};
		\end{loglogaxis}
		\end{tikzpicture}
	}
	\hfill
	\subfloat[error $E_\infty$]{
		\begin{tikzpicture}
		\begin{loglogaxis}[enlargelimits=false,xmin=500,xmax=2100,ymin=1e-15,ymax=1e-3,height=0.33\textwidth, width=0.43\textwidth, grid=major, xlabel={$n_i$}, ylabel={$E_\infty$},
		xtick={514,1024,2048}, log x ticks with f point, 
		ytick={1e-4,1e-10,1e-14},
		legend style={at={(0.5,1.07)}, anchor=south,legend columns=3,legend cell align=left, font=\small, 
		},]
		\addplot[red,mark=o,mark size=2] coordinates {
			(514,8.633164822255256e-5) (614,1.5517866587005866e-11) (716,5.277983613797277e-14) (820,3.7886924134980585e-15) (922,1.608327243190527e-15) (1024,2.446926455759619e-15) (1126,2.05715308667823e-15) (1228,1.8155788579308285e-15) (1332,2.525798196998203e-15) (1434,3.788825115524145e-15) (1536,2.6837632510052974e-15) (1638,2.525821167066744e-15) (1740,2.052208500946621e-15) (1844,2.999381682997571e-15) (1946,1.8954380512017872e-15) (2048,2.2101129544135255e-15) 
		};
		\end{loglogaxis}
		\end{tikzpicture}
	}
	\caption{Errors for one dimensional NFFT and polynomial $f_1$, see \eqref{eq:trigpolys}, with different FFT lengths $\b n$ and fixed window size $m = 8$.} 
	\label{fig:fftw_error:1}
\end{figure}
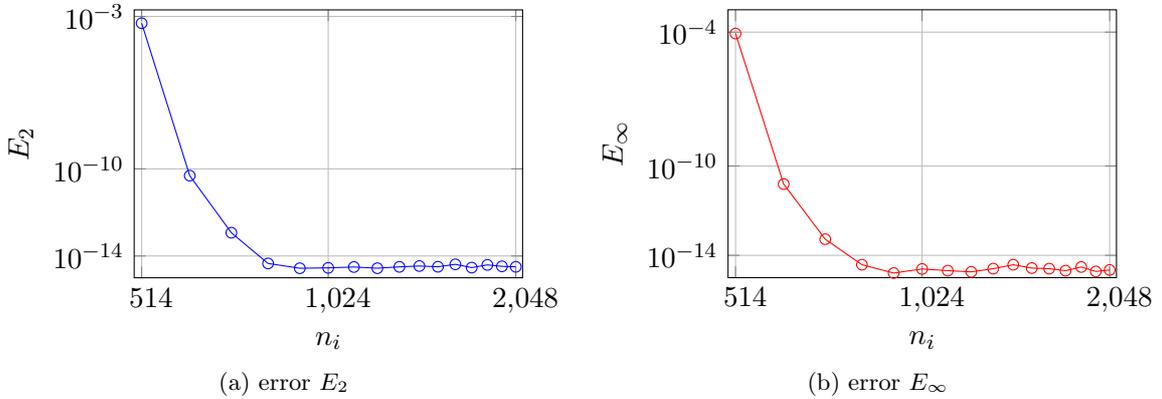
\begin{figure}[ht!]
	\subfloat[error $E_2$]{
		\begin{tikzpicture}
		\begin{loglogaxis}[enlargelimits=false,xmin=126,xmax=530,ymin=1e-15,ymax=1e-2,height=0.33\textwidth, width=0.43\textwidth, grid=major, xlabel={$n_i$}, ylabel={$E_2$},
		xtick={130,256,512}, log x ticks with fixed point, 
		ytick={1e-6,1e-3,1e-10,1e-14},
		legend style={at={(0.5,1.07)}, anchor=south,legend columns=3,legend cell align=left, font=\small, 
		},]
		\addplot[blue,mark=o,mark size=2] coordinates {
			(130,0.0017340046869389937) (154,7.782350944545588e-10) (180,1.6267557585267359e-12) (204,4.874049449025143e-14) (230,8.300475175396248e-15) (256,8.810377336544599e-15) (282,7.728830066375853e-15) (308,7.620557444653936e-15) (332,7.022334199332407e-15) (358,7.539039597491701e-15) (384,7.564924394741687e-15) (410,7.494683554624721e-15) (436,7.798459552881457e-15) (460,7.361770995393079e-15) (486,7.475660306911156e-15) (512,7.334349327437027e-15)
		};
		\end{loglogaxis}
		\end{tikzpicture}
	}
	\hfill
	\subfloat[error $E_\infty$]{
		\begin{tikzpicture}
		\begin{loglogaxis}[enlargelimits=false,xmin=126,xmax=530,ymin=1e-15,ymax=1e-2,height=0.33\textwidth, width=0.43\textwidth, grid=major, xlabel={$n_i$}, ylabel={$E_\infty$},
		xtick={130,256,512}, log x ticks with fixed point, 
		ytick={1e-3,1e-6,1e-10,1e-14},
		legend style={at={(0.5,1.07)}, anchor=south,legend columns=3,legend cell align=left, font=\small, 
		},]
		\addplot[red,mark=o,mark size=2] coordinates {
			(130,0.00010195900713114435) (154,7.737733233829851e-11) (180,1.6591039341802999e-13) (204,8.046358853365855e-15) (230,2.0791989356122773e-15) (256,6.427172378307811e-15) (282,2.4778949209141273e-15) (308,2.811688849465145e-15) (332,1.5403243136038503e-15) (358,3.41437292282206e-15) (384,2.8125186193037763e-15) (410,3.280394043775822e-15) (436,2.946979129703741e-15) (460,2.6108496243625904e-15) (486,2.008615973750585e-15) (512,2.8116926665924996e-15) 
		};
		\end{loglogaxis}
		\end{tikzpicture}
	}
	\caption{Errors for two dimensional NFFT and polynomial $f_2$, see \eqref{eq:trigpolys}, with different FFT lengths $\b n$ and fixed window size $m = 8$.} 
	\label{fig:fftw_error:2}
\end{figure}
\begin{figure}[ht!]
	\subfloat[error $E_2$]{
		\begin{tikzpicture}
		\begin{loglogaxis}[enlargelimits=false,xmin=33,xmax=133,ymin=1e-15,ymax=3e-1,height=0.33\textwidth, width=0.43\textwidth, grid=major, xlabel={$n$}, ylabel={$E_2$},
		xtick={34,128,64}, log x ticks with fixed point, 
		ytick={1e-6,1e-1,1e-10,1e-14},
		legend style={at={(0.5,1.07)}, anchor=south,legend columns=3,legend cell align=left, font=\small, 
		},]
		\addplot[blue,mark=o,mark size=2] coordinates {
			(34,0.05222800449020191) (38,1.8962975233778884e-8) (44,1.1002460800255678e-11) (52,8.945983021549152e-14) (58,1.2892225239854013e-14) (64,1.059318133505856e-14) (70,8.294568008511274e-15) (76,8.325679269613268e-15) (84,7.950432498164398e-15) (90,6.715816287055029e-15) (96,1.0116760849761688e-14) (102,9.828804758155922e-15) (108,6.554701745829851e-15) (116,7.513042200871556e-15) (122,8.129973751487961e-15) (128,9.519306489942944e-15) 
		};
		\end{loglogaxis}
		\end{tikzpicture}
	}
	\hfill
	\subfloat[error $E_\infty$]{
		\begin{tikzpicture}
		\begin{loglogaxis}[enlargelimits=false,xmin=33,xmax=133,ymin=1e-15,ymax=1e-2,height=0.33\textwidth, width=0.43\textwidth, grid=major, xlabel={$n$}, ylabel={$E_\infty$},
		xtick={34,128,64}, log x ticks with fixed point, 
		ytick={1e-3,1e-6,1e-10,1e-14},
		legend style={at={(0.5,1.07)}, anchor=south,legend columns=3,legend cell align=left, font=\small, 
		},]
		\addplot[red,mark=o,mark size=2] coordinates {
			(34,0.001143887265350601) (38,1.0885534869131716e-9) (44,1.4604614903041304e-12) (52,1.5181112212248668e-14) (58,5.833034433030104e-15) (64,6.855992515675157e-15) (70,4.2902140099194976e-15) (76,5.616221765305323e-15) (84,3.630436737343738e-15) (90,3.274444503461071e-15) (96,8.97802163902648e-15) (102,8.977593727297586e-15) (108,3.0662405653202272e-15) (116,4.388711103672799e-15) (122,3.060839697913325e-15) (128,9.79428079312225e-15) 
		};
		\end{loglogaxis}
		\end{tikzpicture}
	}
	\caption{Errors for three dimensional NFFT and polynomial $f_3$, see \eqref{eq:trigpolys}, with different FFT lengths $\b n$ and fixed window size $m = 8$.} 
	\label{fig:fftw_error:3}
\end{figure}

Figures \ref{fig:fftw_error:1}, \ref{fig:fftw_error:2} and \ref{fig:fftw_error:3} show that both errors $E_2$ (see \eqref{eq:E2}) and $E_\infty$ (see \eqref{eq:EI}) are monotonically decreasing with increasing FFT length $\b n$ and stagnation after we reach about $2\b N$. We can observe the same behavior for all three polynomials. 

A detailed analysis of the different parts of the NFFT in \cite{Vo12} shows that the FFT takes only a small portion of time compared to the other parts. The effect is even more prominent as the dimension increases. This indicates that lowering $\b n$ can cause a large deterioration in the error while time improvements might be marginal.

\subsubsection{Window Size $m$}\label{sec:window}

The window size $m$ determines the size of the index sets $I_{\b n,m}(\b{x}_j)$ (see \eqref{eq:Inm}) and will therefore have a large expected impact on both accuracy of the results and computation times. The standard value for this parameter is $m = 8$. Note that this is only true for the standard window function and results for $m$ may vary for other windows. For our tests, we will look at all values $m = 2,3,\dots,10$.

\begin{figure}[ht!]
	\subfloat[error $E_2$]{
		\begin{tikzpicture}
		\begin{loglogaxis}[enlargelimits=false,xmin=1.9,xmax=11,ymin=1e-15,ymax=1e-3,height=0.33\textwidth, width=0.43\textwidth, grid=major, xlabel={$n$}, ylabel={$E_2$},
		xtick={2,3,5,8,10}, log x ticks with fixed point,
		ytick={1e-4,1e-6,1e-10,1e-14},
		legend style={at={(0.5,1.07)}, anchor=south,legend columns=3,legend cell align=left, font=\small, 
		},]
		\addplot[blue,mark=o,mark size=2] coordinates {
			(2,5.563266967355224e-5) (3,6.013580418630189e-7) (4,1.633919357784835e-8) (5,1.151589459152576e-10) (6,9.67165786908648e-13) (7,1.6127912908794474e-14) (8,2.8505654409202513e-15) (9,4.078971742073087e-15) (10,4.2108197356584394e-15) 
		};
		\addlegendentry{$d$=1}
		\addplot[red,mark=square,mark size=1.5] coordinates {
			(2,0.0002519679617877262) (3,2.713643650756638e-6) (4,3.19769871613658e-8) (5,3.472903178388193e-10) (6,3.655973417292403e-12) (7,4.408261558514131e-14) (8,8.810377336544599e-15) (9,9.252599024788408e-15) (10,1.0557511231625159e-14) 
		};
		\addlegendentry{$d$=2}
		\addplot[green!75!black,mark=triangle,mark size=2.5] coordinates {
			(2,0.00037966806393413674) (3,4.508641107680969e-6) (4,5.95188521061288e-8) (5,7.126289425973831e-10) (6,8.37329295124224e-12) (7,1.0271659110224542e-13) (8,1.059318133505856e-14) (9,1.0643394276963766e-14) (10,1.2002643126010844e-14) 
		};
		\addlegendentry{$d$=3}
		\end{loglogaxis}
		\end{tikzpicture}
	}
	\hfill
	\subfloat[error $E_\infty$]{
		\begin{tikzpicture}
		\begin{loglogaxis}[enlargelimits=false,xmin=1.9,xmax=11,ymin=1e-15,ymax=1e-3,height=0.33\textwidth, width=0.43\textwidth, grid=major, xlabel={$n$}, ylabel={$E_\infty$},
		xtick={2,3,5,8,10}, log x ticks with fixed point,
		ytick={1e-4,1e-6,1e-10,1e-14},
		legend style={at={(0.5,1.07)}, anchor=south,legend columns=3,legend cell align=left, font=\small, 
		},]
		\addplot[blue,mark=o,mark size=2] coordinates {
			(2,5.563266967355224e-5) (3,6.013580418630189e-7) (4,1.633919357784835e-8) (5,1.151589459152576e-10) (6,9.67165786908648e-13) (7,1.6127912908794474e-14) (8,2.8505654409202513e-15) (9,4.078971742073087e-15) (10,4.2108197356584394e-15) 
		};
		\addlegendentry{$d$=1}
		\addplot[red,mark=square,mark size=1.5] coordinates {
			(2,0.00012104760827902332) (3,1.2091466458185785e-6) (4,1.09343311083641e-8) (5,1.2243888637685437e-10) (6,9.93800902539681e-13) (7,1.0141543216969679e-14) (8,6.427172378307811e-15) (9,4.351438604287648e-15) (10,4.6862347985158185e-15)  
		};
		\addlegendentry{$d$=2}
		\addplot[green!75!black,mark=triangle,mark size=2.5] coordinates {
			(2,0.0001023555328755899) (3,8.676174917988886e-7) (4,1.2224628694477626e-8) (5,1.1299327817668467e-10) (6,1.7689586542627788e-12) (7,2.0257914398115042e-14) (8,6.855992515675157e-15) (9,6.835338400907806e-15) (10,5.332177746503094e-15) 
		};
		\addlegendentry{$d$=3}
		\end{loglogaxis}
		\end{tikzpicture}
	}
	\caption{Errors for one, two and three dimensional NFFT with polynomials $f_d$, see \eqref{eq:trigpolys}, different window size $m$ and fixed FFT length $\b n = 2\b N$.} 
	\label{fig:window_size_error}
\end{figure}
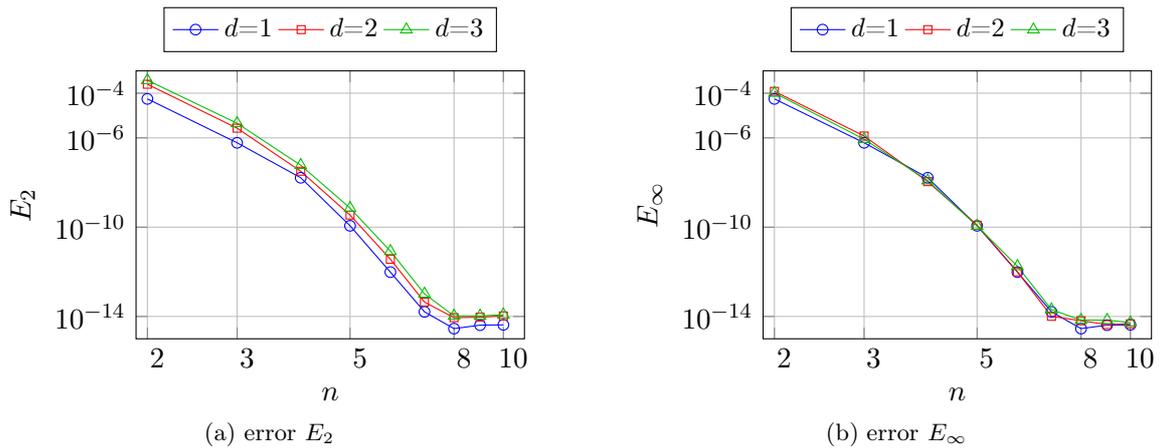
\begin{figure}[ht!]
	\subfloat[$d = 1$]{
		\begin{tikzpicture}
		\begin{loglogaxis}[enlargelimits=false,xmin=1.9,xmax=10.5,ymin=0.03,ymax=0.9,height=0.33\textwidth, width=0.43\textwidth, grid=major, xlabel={$m$}, ylabel={time in ms},
		xtick={2,3,5,8,10}, log x ticks with fixed point,
		ytick={0.04,0.06,0.3,0.85,0.5}, log y ticks with fixed point,
		legend style={at={(0.5,1.07)}, anchor=south,legend columns=3,legend cell align=left, font=\small, 
		},]
		\addplot[blue,mark=o,mark size=2] coordinates {
			(2,0.2837138) (3,0.3828264) (4,0.4581253) (5,0.5389944) (6,0.563706) (7,0.645324) (8,0.685525) (9,0.7454571999999999) (10,0.8064241999999999)  
		};
		\addlegendentry{precompute time}
		\addplot[red,mark=square,mark size=1.5] coordinates {
			(2,0.039756400000000004) (3,0.043083300000000005) (4,0.04509) (5,0.0584921) (6,0.0500644) (7,0.0548865) (8,0.056452800000000004) (9,0.0570821) (10,0.0595809) 
		};
		\addlegendentry{trafo time}
		\end{loglogaxis}
		\end{tikzpicture}
	}
	\hfill
	\subfloat[$d = 2$]{
		\begin{tikzpicture}
		\begin{loglogaxis}[enlargelimits=false,xmin=1.9,xmax=10.5,ymin=7.5,ymax=65,height=0.33\textwidth, width=0.43\textwidth, grid=major, xlabel={$m$}, 
		xtick={2,3,5,8,10}, log x ticks with fixed point,
		ytick={8,20,35,55,15}, log y ticks with fixed point,
		legend style={at={(0.5,1.07)}, anchor=south,legend columns=3,legend cell align=left, font=\small, 
		},]
		\addplot[blue,mark=o,mark size=2] coordinates {
			(2,20.0698369) (3,25.6931681) (4,33.5185778) (5,35.514404) (6,39.3356485) (7,41.4693532) (8,45.587228100000004) (9,49.5665897) (10,53.553691)
		};
		\addlegendentry{precompute time}
		\addplot[red,mark=square,mark size=1.5] coordinates {
			(2,8.1391221) (3,8.277465900000001) (4,10.7805905) (5,12.9487957) (6,16.1711527) (7,18.181554100000003) (8,21.3089698) (9,24.8293761) (10,36.0697724) 
		};
		\addlegendentry{trafo time}
		\end{loglogaxis}
		\end{tikzpicture}
	}
	\par\medskip
	\centering
	\subfloat[$d = 3$]{
		\begin{tikzpicture}
		\begin{loglogaxis}[enlargelimits=false,xmin=1.9,xmax=10.5,ymin=36,ymax=1700,height=0.33\textwidth, width=0.43\textwidth, grid=major, xlabel={$m$}, 
		xtick={2,3,5,8,10}, log x ticks with fixed point,
		ytick={50,200,500,1500}, log y ticks with fixed point,
		legend style={at={(0.5,1.07)}, anchor=south,legend columns=3,legend cell align=left, font=\small, 
		},]
		\addplot[blue,mark=o,mark size=2] coordinates {
			(2,60.0010256) (3,76.8782115) (4,93.50766340000001) (5,109.08808379999999) (6,114.4381968) (7,124.8921232) (8,138.3052241) (9,153.4258524) (10,163.4729829)
		};
		\addlegendentry{precompute time}
		\addplot[red,mark=square,mark size=1.5] coordinates {
			(2,39.5714381) (3,80.4174904) (4,191.9019791) (5,311.1908948) (6,468.0651797) (7,657.491096375) (8,896.8205791666667) (9,1182.301242) (10,1529.23583875) 
		};
		\addlegendentry{trafo time}
		\end{loglogaxis}
		\end{tikzpicture}
	}
	\caption{Computation times for one, two and three dimensional NFFT with polynomials $f_d$, see \eqref{eq:trigpolys}, different window size $m$ and fixed FFT length $\b n = 2\b N$.} 
	\label{fig:window_size_times} 
\end{figure}
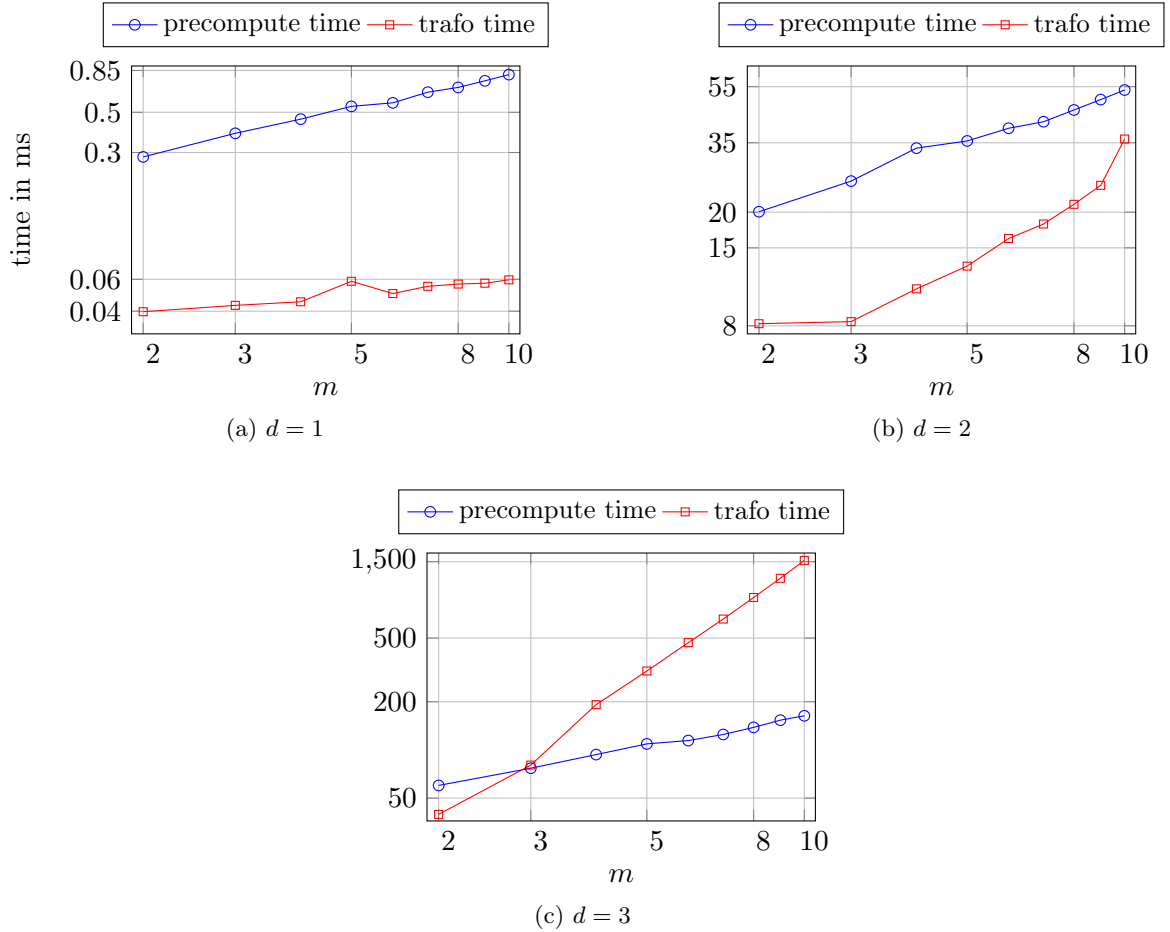

In Figure \ref{fig:window_size_error}, we can see that both errors decrease monotonically until a window size of $m=8$. After that we have stagnation at machine precision (for this example with the standard window function). Therefore, it makes sense to choose $m$ between $2$ and $8$ according to the desired accuracy. 

Figure \ref{fig:window_size_times} shows that the window size has a significant impact on both the precomputation and transformation times. We can observe an increase with a factor of about $3$ for the precompute time between $m=2$ and $m=10$ for all polynomials. The actual transformation takes about $1.5$ times longer for $d=1$, $4.43$ times for $d=2$ and even $38.6$ times as long in the case of $d=3$. 

\subsection{Impact of OpenMP}\label{sec:openmp}

Now, we want to take a look at the impact of OpenMP on the NFFT computation times. In order to have a high enough required base time to keep the influence of noise minimal, we take the polynomial $f_3$ from \eqref{eq:trigpolys} with multibandlimit $\b N = (128,128,128)$ and one node set with $M = 2\cdot {256}^3$ nodes.

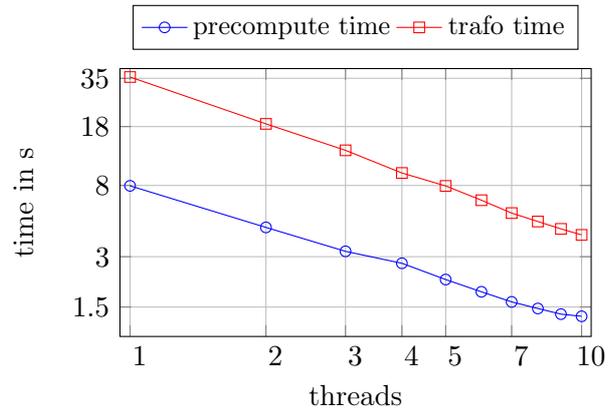
\begin{figure}[ht!]
	\begin{center}
		\begin{tikzpicture}
		\begin{loglogaxis}[enlargelimits=false,xmin=9.5e-1,xmax=10.5,ymin=1,ymax=40,height=0.33\textwidth, width=0.5\textwidth, grid=major, xlabel={threads}, ylabel={time in s},
		xtick={1,2,5,4,3,7,10}, log x ticks with fixed point, 
		ytick={35,18,8,1.5,3}, log y ticks with fixed point,
		legend style={at={(0.5,1.07)}, anchor=south,legend columns=3,legend cell align=left, font=\small, 
		},]
		\addplot[blue,mark=o,mark size=2] coordinates {
			(1,7.96) (2,4.49) (3,3.23) (4,2.74) (5,2.19) (6,1.85) (7,1.61) (8,1.47) (9,1.36) (10,1.32)
		};
		\addlegendentry{precompute time}
		\addplot[red,mark=square,mark size=2] coordinates {
			(1,35.62) (2,18.68) (3,12.98) (4,9.50) (5,7.95) (6,6.53) (7,5.47) (8,4.87) (9,4.40) (10,4.05)
		};
		\addlegendentry{trafo time}
		\end{loglogaxis}
		\end{tikzpicture}
	\end{center}
	\caption{Computation times for NFFT with polynomial $f_3$, multibandlimit $\b N = (128,128,128)$, $M = 2\cdot 128^3$, $\b n = 2\b N$, and $m = 6$ using OpenMP.} 
	\label{fig:omp}
\end{figure}

Figure \ref{fig:omp} shows that we have a decrease in time for precomputations and transformation as we raise the number of OMP threads. The threads are limited by the capabilities of the processor and it does not make sense to pick them higher as the number of available cores. If one has to calculate multiple NFFTs, possibly at the same time, it can be beneficial to look into both OMP and Multi-Core processing directly in Julia.

\end{document}